\documentclass[11pt]{article}
\pdfoutput=1
\usepackage[latin1]{inputenc}
\usepackage{amsmath}
\usepackage{amsfonts}
\usepackage{amssymb}
\usepackage{amsthm}
\usepackage{mathrsfs}
\usepackage{amssymb}
\usepackage{cancel}
\usepackage{slashed}
\usepackage{graphicx}
\usepackage[font=small,labelfont=bf]{caption}
\usepackage{float}
\usepackage{csquotes}
\usepackage[bottom]{footmisc} 
\usepackage{tikz}
\usepackage{setspace}
\usepackage{bbding}
\usepackage{cite}
\usepackage{url}
\urldef\myurl\path{https://en.wikipedia.org/wiki/List_of_formulas_in_Riemannian_geometry#Ricci_tensor} 
\usepackage{array,longtable}
\makeatletter

\@addtoreset{equation}{section}
\makeatother

\def\lsim{\;\raise0.3ex\hbox{$<$\kern-0.75em\raise-1.1ex\hbox{$\sim$}}\;}
\def\gsim{\;\raise0.3ex\hbox{$>$\kern-0.75em\raise-1.1ex\hbox{$\sim$}}\;}
\def\beq{\begin{equation}}   \def\eeq{\end{equation}}
\def\ba{\begin{array}}       \def\ea{\end{array}}
\def\bea{\begin{eqnarray}}   \def\eea{\end{eqnarray}}
\def\nn{\nonumber}
\def\nl{\newline}

\theoremstyle{definition} 
\date{\today}

\begin{document}

\begin{titlepage}

\begin{center}


\vspace{1cm}
{\Large\bf Gradient Flow and Holography from a Local Wilsonian Cutoff} 
\vspace{2cm}

{\bf{Ulrich Ellwanger}}\\
\vspace{1cm}
{\it  University  Paris-Saclay,  CNRS/IN2P3,  IJCLab,  91405  Orsay,  France}

\end{center}

\vspace*{2cm}
\begin{abstract}

We consider the vacuum partition function of a 4d scalar QFT in a curved background as function of bare marginal and relevant couplings. A local UV cutoff $\Lambda(x)$ transforming under Weyl rescalings allows to construct Weyl invariant kinetic terms including Wilsonian cutoff functions. The local cutoff can be absorbed completely by a rescaling of the metric and the bare couplings. The vacuum partition function satisfies consistency conditions which follow from the Abelian nature of local redefinitions of the cutoff, and which differ from Weyl rescalings. These imply a gradient flow for beta functions describing the cutoff dependence of rescaled bare couplings. The consistency conditions allow to satisfy all but one Hamiltonian constraints required for a holographic description of the flow of bare couplings with the cutoff.

\end{abstract}

\end{titlepage}

\section{Introduction}

In quantum field theories (QFTs) coupling constants including masses can be promoted to local functions of space-time in which case they become sources for corresponding operators. If QFTs are considered in curved space-time, local Weyl transformations can be defined. These are sensitive to the ultraviolet regularization and renormalization of QFTs, i.e. to counter terms and beta functions. The Abelian nature of local Weyl transformations implies consistency conditions on coefficients of the vacuum partition function and beta functions. This program was initiated by Osborn and Jack and Osborn [JO] in \cite{Osborn:1989td,Jack:1990eb,Osborn:1991gm}.
Implied relations among coefficients of $\beta$~functions in dimensional regularization have been studied in \cite{Antipin:2013sga,Jack:2013sha,Baume:2014rla,Jack:2014pua,Jack:2015tka,Antipin:2018brm,Poole:2019txl,Poole:2019kcm,Davies:2019onf,Sartore:2020pkk,Steudtner:2020tzo} for the Standard Model and others.

In dimensional regularization classical Weyl invariance is broken by the scale $\mu$ introduced for otherwise dimensionless couplings. Here we consider a regularization by a Wilsonian cutoff $\Lambda$, by which we understand a modification of kinetic terms in the action such that modes with momenta $p^2 \gg \Lambda^2$ are suppressed, i.e. the kinetic terms become very (e.g. exponentially) large for $p^2 \gg \Lambda^2$. We consider a local cutoff $\Lambda(x)$ transforming under local Weyl transformations which allows to define local renormalization group equations (RGEs). A local RGE was considered as early as 1987 in \cite{Shore:1986hk} in two-dimensional curved spacetime sigma models in order to derive consistency conditions on allowed backgrounds in string theory. A local Wilsonian cutoff was considered e.g. in \cite{Codello:2012sn,Codello:2015ana}, with the aim to study exact (functional) RGEs in 2 and 4 dimensions.

In the present paper we focus on the dependence of bare couplings on the cutoff, and on the bare vacuum partition function before adding counter terms. The dependence of bare couplings on the cutoff is also described by beta functions, which are functions of bare couplings and of the cutoff in case of relevant couplings (such as masses) as considered here. Via subdivergences, the bare vacuum partition function is also sensitive to the cutoff dependence of bare couplings. 
(A Wilsonian cutoff underlines the hierarchy problem, i.e. renormalized and bare masses are naturally of the order of the UV cutoff. For renormalized masses this might be solved by fine tuning or by Supersymmetry; the subsequent results do not depend on whether or how this issue is solved.\footnote{We remark that we will not employ the (exact) Wilsonian renormalization group.})

The introduction of a local Wilsonian cutoff transforming under Weyl transformations allows to construct Weyl invariant kinetic terms for scalars in 4d \cite{Ellwanger:2021orv}; here the local component of the cutoff plays the role of a compensator. 
Assuming massive fields only, one finds that the local cutoff dependence of the bare vacuum partition function can be absorbed completely by rescalings of the space-time metric and bare relevant couplings, a particular feature of a Wilsonian cutoff. This allows to express the dependence of the bare vacuum partition function on the local cutoff in terms of its dependence on the rescaled space-time metric and, via corresponding beta functions, on bare marginal and rescaled relevant couplings. 

The possibility to absorb the local cutoff by rescalings of the space-time metric and bare relevant couplings does not imply the absence of anomalies. Assuming massive fields only, the cutoff appears in the form of logarithms $\log(\Lambda^2/m_0^2)$, i.e. bare masses play the role of the renormalization scale $\mu$ in dimensional regularization. After a rescaling, bare masses squared $m_0^2$ become $\Lambda^2 \hat{m}_0^2$ with $\hat{m}_0^2$ dimensionless, hence logarithms $\log(\Lambda^2/m_0^2)$ become $-\log(\hat{m}_0^2)$. In the bare vacuum partition function such logarithms, multiplying e.g. the Euler density which is not Weyl invariant, indicate the presence of Weyl anomalies as studied (including their dependence on dimensionless couplings) by [JO].

Local transformations of the cutoff are Abelian transformations similar to (but different from) local Weyl transformations. Accordingly they lead to consistency conditions similar to (but different from) local Weyl transformations. These consistency conditions imply a gradient flow for the beta functions describing the cutoff dependence of bare marginal and rescaled relevant couplings. In an expansion to first non-trivial order in derivatives 
the role of the potential of the gradient flow is played by the coefficient of the Riemann scalar $R$ in the bare vacuum partition function.

Subsequently we turn to a possible holographic description of the flow of bare couplings with the cutoff.
Soon after the birth of the AdS/CFT correspondence it became clear that a cutoff on the extra coordinate $\rho$ for $\rho \to 0$ of an asymptotic AdS space can be related to a UV cutoff or a RG scale of a conformal field theory (CFT) (see \cite{Skenderis:2002wp} for an early review). A finite result for the CFT action on the boundary is obtained only if the induced metric and the fields are rescaled by appropriate powers of $\rho$. In addition, extra counter terms are required to cancel logarithmic divergences multiplying e.g. the Euler density. Requiring these counter terms to be local and covariant, Henningson and Skenderis have shown in \cite{Henningson:1998gx} that the so obtained finite action on the boundary still varies under conformal transformations of the boundary metric, and that the corresponding Weyl anomalies match the ones of the CFT. In the present approach counter terms are omitted, and the Weyl anomalies appear directly as non-invariance under conformal transformations of corresponding terms in the bare action on the boundary. However, within our expansion to second order in derivatives these Weyl anomalies are not yet visible for a 4d QFT on the boundary of a 5d bulk for which Weyl anomalies are of fourth order in derivatives.

It was frequently underlined, however, that a better understanding of the precise relation between a UV cutoff on a QFT and on the extra coordinate of an asymptotic AdS space would be desirable (see e.g. \cite{Heemskerk:2010hk,Nakayama:2015ita,Hartman:2018tkw}). It is clear that dimensional regularization of a QFT is of little help here as the notion of a UV cutoff does not exist.

In the framework of classical general relativity in a 5d bulk, the flow with the fifth coordinate of fields living on a 4d boundary can be described within the Hamilton-Jacobi formalism \cite{deBoer:1999tgo} (see also \cite{Fukuma:2002sb}). The Hamiltonian constraint implies constraints on the coefficients of the action on the 4d boundary, and on the beta functions once the fields on the boundary are identified with running bare couplings of a 4d QFT. For these beta functions one of the Hamiltonian constraints implies a gradient flow. Based on this feature a condition for a holographic duality was proposed in \cite{Balasubramanian:2000wv}. (An interpretation of the renormalization group flow as a Hamiltonian vector flow was already proposed by B.~Dolan in 1994 \cite{Dolan:1994eg}.)

Only few studies concern the question whether the RG flows of generic QFTs (non-CFTs) admit a holographic description \cite{Kiritsis:2014kua}. 
Confining oneself to classical general relativity in the 5d bulk, mostly two approaches were persued in order to establish a link between 4d QFTs (conformal or non-CFTs) and holography: One approach uses the formal similarity between the Hamilton-Jacobi formalism (quadratic in functional derivatives of the action on the boundary with respect to fields) and Wilson's (or Polchinski's \cite{Polchinski:1983gv}) exact RG, see e.g. \cite{Li:2000ec,Akhmedov:2010sw,Leigh:2014qca,Lizana:2015hqb}. 
Alternatively one can focus on the link between Weyl transformations acting on the 4d QFT/CFT vacuum partition function on the boundary and diffeomorphisms in an AdS$_5$ bulk \cite{fefferman}, see e.g. \cite{Imbimbo:1999bj,Erdmenger:2001ja,Nakayama:2013is,Rajagopal:2015lpa,Kikuchi:2015hcm,Shyam:2017qlr} for corresponding studies.
An approach towards quantum gravity in the bulk from a matrix field theory via quantum RG has been persued in \cite{Lee:2013dln}.

A known issue is whether a description of a QFT on an (inner) boundary of AdS in terms of a local Lagrangian in the bulk is possible at all \cite{Banks:1998dd} (see also \cite{El-Showk:2011yvt} and refs. therein). Here we assume this to be the case, which corresponds to the assumption that an expansion of a local covariant 5d Lagrangian in derivatives exists allowing to obtain connected N point functions of fields at different points. For points on the boundary, these must match the N point functions of the QFT after suitable rescalings.

Here we address the question whether the RG flow of bare marginal and relevant couplings of a generic massive (non-CFT) scalar QFT admits a holographic dual. The number of bare marginal and relevant couplings will be denoted by $n_g$.
We review shortly the Hamilton-Jacobi formalism leading to flow equations for fields and the Hamiltonian constraint on the action on the boundary of a 5d bulk. Given a 5d Lagrangian in the bulk, the Hamiltonian constraint leads to constraints on coefficients of the action on the boundary. Assuming this action to be given by the vacuum partition function of a generic QFT expanded to second order in derivatives, its coefficients seem to be heavily over-determined.

However, we find that this is not so if the following dictionnary is applied:\\
1) The fields on the boundary of the 5d bulk are identified with 4d bare marginal or rescaled relevant couplings, and the boundary action with the 4d vacuum partition function.\\
2) The 4d metric on the boundary of the 5d bulk satisfies a field dependent flow equation. Nevertheless, the 4d metric on the boundary of the 5d bulk is identified with the 4d QFT metric after its rescaling by the local cutoff $\Lambda(x)$. This implies that, implicitely, the fifth bulk coordinate on the boundary is related in a field dependent way to the local QFT cutoff $\Lambda$. (As expected, the 5d bulk metric is AdS up to field dependent corrections.) \\
3) Then $n_g$ consistency conditions on coefficients of the QFT originating from the Abelian nature of local cutoff transformations coincide exactly with $n_g$ Hamiltonian constraints which seem to over-determine the vacuum partition function of a generic QFT. The remaining Hamiltonian constraints allow to reconstruct a suitable 5d bulk Lagrangian for a generic (massive) 4d QFT provided one additional condition
(within the considered order of an expansion in derivatives) is satisfied. This is the main result of the second part of this paper.

In Section~2 we construct Weyl invariant kinetic terms including a Wilsonian cutoff as well as interaction terms following \cite{Ellwanger:2021orv}. We introduce a rescaled metric $\hat\gamma_{\mu\nu}$ and rescaled bare couplings $\hat g_0^a$ in terms of which the explicit dependence of the vacuum partition function $\hat S_{QFT}$ on the local cutoff disappears. A consistency condition to be satisfied by $\hat S_{QFT}$, expanded to second order in derivatives, following from the Abelian nature of local variations of the cutoff is derived. This consistency condition assumes the form of an exact gradient flow for the beta functions.

In Section~3 we review the Hamilton-Jacobi formalism following to a large extend refs.~\cite{deBoer:1999tgo,Fukuma:2002sb}. Parts of the derivation are given in Appendix~A. 
The corresponding flow equations and Hamiltonian constraints are derived. In Section~4 we show that, within the expansion up to second order in derivatives, all Hamiltonian constraints up to one can be satisfied by an arbitrary 4d scalar QFT thanks to the consistency condition following from the local Wilsonian cutoff. A summary and outlook is given in Section~5.

\section{Gradient Flow From Weyl Invariant Wilsonian Cutoff Functions}

In this Section we construct a Weyl invariant kinetic term, including a Wilsonian UV cutoff, for a real scalar field $\varphi$ in $d$ dimensions. The UV cutoff $\Lambda$ is assumed to be local, i.e. $\Lambda(x)$, and to transform under Weyl rescalings together with the background metric $\gamma_{\mu\nu}$ and the scalar field:
\bea
&\delta_\sigma \gamma_{\mu\nu}=& -2\sigma \gamma_{\mu\nu}\; ,
\nn \\
&\delta_\sigma \varphi=&  \sigma \left(\frac{d}{2}-1\right) \varphi\; ,
\nn \\
&\delta_\sigma \Lambda=& \sigma \Lambda\; .
\label{eq:weylresc}
\eea
One may write the local cutoff $\Lambda(x)$ in the form
\beq\label{Ltol}
\Lambda(x)=\overline{\Lambda} e^{\lambda(x)}\; ,\quad \delta_\sigma \overline{\Lambda} =0\; ,\quad \delta_\sigma \lambda(x) = \sigma(x) \; .
\eeq
$\lambda(x)$ plays the role of a compensator: Due to the presence of $\lambda(x)$ the effective action satisfies a local RGE (see eq.~\eqref{localRGEW0} below), but in the ``gauge'' $\lambda(x)=0$ this hold no longer.

For the construction of a kinetic term including a local Wilsonian cutoff we start with the known expression for a Weyl covariant generalization of the covariant Laplacian
\beq
\nabla^2-\xi R\; , \quad 
\nabla^2=\frac{1}{\sqrt{\gamma}}\partial_\mu\sqrt{\gamma}\gamma^{\mu\nu}\partial_\nu\; ,
\quad \gamma=\det({\gamma_{\mu\nu}})\; ,
\quad \xi=\frac{d-2}{4(d-1)}\; ,
\eeq
which satisfies
\beq
\delta_\sigma\left[(\nabla^2-\xi R){\cal O}\right] = \left(\frac{d}{2}+1\right) \sigma \left[(\nabla^2-\xi R){\cal O}\right]
\eeq
provided the operator $\cal{O}$ satisfies $\delta_\sigma {\cal O}=\sigma \left(\frac{d}{2}-1\right) {\cal O}$ such that, for ${\cal O}=\varphi$, $\delta_\sigma (\sqrt{\gamma}\varphi(\nabla^2-\xi R)\varphi)=0$ (using $\delta_\sigma \sqrt{\gamma}=-d\,\sigma \sqrt{\gamma}$). A Weyl invariant Wilsonian cutoff can be constructed with help of the operator
\beq
D_\Lambda = \Lambda^{-2}(\nabla^2-\xi R)\; .
\eeq
Using $\delta_\sigma \Lambda^{-2}=-2\sigma \Lambda^{-2}$, any expression of the form 
\beq\label{eq:FD}
F(D_\Lambda)
\eeq
satisfies
\beq
\delta_\sigma \left( F(D_\Lambda) \varphi\right) = \left(\frac{d}{2}-1\right)\sigma  F(D_\Lambda) \varphi\; .
\eeq
The possibility to construct $F(D_\Lambda)$ with this property under local Weyl transformations requires the use of the $x$~dependent cutoff $\Lambda$ transforming as in eq.~\eqref{eq:weylresc}. An example is
\beq\label{eq:FDe}
F(D_\Lambda)=e^{-D_\Lambda}
\eeq
which leads to an exponentially suppressed propagator for large $p^2$ in momentum space.
The kinetic part $S_k$ of the action reads then
\beq\label{eq:kin}
S_k(\gamma,\varphi,\Lambda)=\frac{1}{2}\int\sqrt{\gamma}d^dx\, \varphi(-\nabla^2+\xi R)\, F(D_\Lambda) \varphi
\eeq
and satisfies
\beq\label{eq:rge1}
\sigma(x)
\left\{-2 \gamma_{\mu\nu}(x)\frac{\delta}{\delta \gamma_{\mu\nu}(x)}
+ \varphi(x) \frac{\delta}{\delta \varphi(x)}
+\Lambda(x) \frac{\delta}{\delta \Lambda(x)}
 \right\}  S_k(\gamma,\varphi,\Lambda)=0\; .
\eeq
In terms of a rescaled metric 
\beq\label{rescgamma}
\hat\gamma_{\mu\nu}=\gamma_{\mu\nu}\Lambda^2
\eeq
eq.\eqref{eq:rge1} becomes
\beq\label{eq:rge2}
\sigma(x)
\left\{\varphi(x) \frac{\delta}{\delta \varphi(x)}
+\Lambda(x) \frac{\delta}{\delta \Lambda(x)}
 \right\}  S_k(\hat\gamma,\varphi,\Lambda)=0\; .
\eeq

As interactions we consider operators ${\cal O}^a(\varphi,\nabla) \equiv {\cal O}^a(\varphi,\gamma_{\mu\nu})$ which satisfy
\beq\label{eq:defop}
\left\{
\int \sqrt{\gamma} d^dz\, \sigma 
\left(-2 \gamma_{\mu\nu}\frac{\delta}{\delta \gamma_{\mu\nu}}
+ \varphi \frac{\delta}{\delta \varphi}\right)
 \right\}\sqrt{\gamma(x)}\, {\cal O}^a(x)= -d^a_0 \sigma(x) \sqrt{\gamma(x)}\, {\cal O}^a(x)\; .
\eeq
A typical case is ${\cal O}^a=\varphi^n$ for which $d^a_0 = d-n$. The interaction part $S_{int}$ involving real scalar fields $\varphi\equiv \left\{\varphi_k\right\}$ (indices of fields will be suppressed for simplicity) reads
\beq\label{eq:sint}
S_{int}(\gamma,\varphi,g_0,\Lambda)=\int \sqrt{\gamma} d^dx \sum_a g^a_0(\Lambda,g)\, {\cal O}^a(x)
\eeq
where $g^a_0(\Lambda,g)$ are bare marginal or relevant couplings of canonical dimension $d^a_0$. They include counter terms depending on $\Lambda$ and on renormalized couplings $g^a$. 
(These counter terms include polynomials in the renormalized couplings and, for relevant couplings, derivatives of $g^a$ as well as of $\gamma$ in the form of suitable contractions of the curvature tensor.) 
In the presence of massless fields $\varphi$ a renormalization scale $\mu$ has to be introduced; otherwise a $\mu$-independent on-shell renormalization scheme can be used.

A Weyl rescaling of $S_{int}(\gamma,\varphi,g_0,\Lambda)$ gives
\begin{align}\label{wresc}
\sigma(x)
\left(-2 \gamma_{\mu\nu}(x)\frac{\delta}{\delta \gamma_{\mu\nu}(x)}
+ \varphi(x) \frac{\delta}{\delta \varphi(x)}
+\Lambda(x) \frac{\delta}{\delta \Lambda(x)}\right)&S_{int}(\gamma,\varphi,g_0,\Lambda)\nn \\
=&-\sigma(x)d_a^0 g^a_0\frac{\delta}{\delta g^a_0} S_{int}(\gamma,\varphi,g_0,\Lambda)\; .
\end{align}
In terms of rescaled bare couplings
\beq\label{rescg}
\hat g_0^a = \Lambda^{-d_0^a} g_0^a
\eeq
and $\hat \gamma_{\mu\nu}$ from \eqref{rescgamma}, $\hat S_{int}(\hat\gamma,\varphi,\hat g_0,\Lambda)$ satisfies
\beq\label{wreschat}
\sigma(x)
\left( \varphi(x) \frac{\delta}{\delta \varphi(x)}
+\Lambda(x) \frac{\delta}{\delta \Lambda(x)}\right) \hat S_{int}(\hat\gamma,\varphi,\hat g_0,\Lambda)
=0\; .
\eeq
The rescaled couplings are still nontrivial functions of renormalized couplings $g$ and the cutoff. After the rescaling the beta functions for relevant couplings (see below) contain tree level terms $\sim -d_0^a$.

The full action $S(\gamma,g_0,\Lambda,\varphi)=S_k+S_{int}$ satisfies a local RGE
\beq\label{rgen0} 
\sigma(x) 
\left\{-2 \gamma_{\mu\nu}(x)\frac{\delta}{\delta \gamma_{\mu\nu}(x)}
+ \varphi(x) \frac{\delta}{\delta \varphi(x)}
+\Lambda(x) \frac{\delta}{\delta \Lambda(x)}
+d_a^0 g^a_0 \frac{\delta}{\delta g^a_0(x)}
 \right\} S(\gamma,g_0,\Lambda,\varphi)  =0\; ,
\eeq
whereas $\hat S(\hat\gamma,\hat g_0,\Lambda,\varphi)$ satisfies
\beq\label{rgen0hat} 
\sigma(x) 
\left\{\varphi(x) \frac{\delta}{\delta \varphi(x)}
+\Lambda(x) \frac{\delta}{\delta \Lambda(x)}
 \right\} \hat S(\hat\gamma,\hat g_0,\Lambda,\varphi)  =0\; .
\eeq

We consider the vacuum partition function $S^0_{QFT}(\gamma,g_0,\Lambda)$ given by
\beq\label{eq:W0}
e^{-{S^0_{QFT}(\gamma,g_0,\Lambda)}} = \frac{1}{\cal N}
\int {\cal D}\varphi\, e^{-S(\gamma,g_0,\Lambda,\varphi)}\; .
\eeq
Since $\varphi$ are dummy variables on the right hand side of eq.~\eqref{eq:W0}, eq.~\eqref{rgen0}
implies a local RGE for ${S^0}_{QFT}(\gamma,g_0,\Lambda)$
\beq\label{localRGEW0}
\sigma(x) 
\left\{-2 \gamma_{\mu\nu}(x)\frac{\delta}{\delta \gamma_{\mu\nu}(x)}
+\Lambda(x) \frac{\delta}{\delta \Lambda(x)}
+d^a_0 g^a_0(x) \frac{\delta}{\delta g^a_0(x)}
 \right\} {S^0}_{QFT}(\gamma,g_0,\Lambda)  =0\; .
\eeq

For ${\Lambda} \to \infty$, the vacuum partition function ${S^0}_{QFT}(\gamma,g_0,\Lambda)$ suffers from subdivergences as well as from superficial divergences. Subdivergences are removed if one expresses the bare couplings $g_0(g,\Lambda)$ explicitely in terms of $\Lambda$ and renormalized cutoff independent couplings $g$, and 
\break 
$S^0_{QFT}(\gamma,g_0(g,\Lambda),\Lambda)$ becomes ${S^0}'_{QFT}(\gamma,g,\Lambda)$. 
The remaining dependence of ${S^0}'_{QFT}(\gamma,g,\Lambda)$ on ${\Lambda}$ concerns superficial divergences which require extra counter terms ${S^{CT}}_{QFT}(\gamma,g,\Lambda,\mu)$; again a renormalization scale $\mu$ is required in the presence of massless fields.
(Anomalies for massless CFTs originate from violations of the local RGE \eqref{localRGEW0} by ${S^{CT}}_{QFT}(\gamma,g,\Lambda,\mu)$ and hence by the finite renormalized vacuum partition function.)
In the following we will not add ${S^{CT}}_{QFT}(\gamma,g,\Lambda,\mu)$ but consider ${S^0}_{QFT}(\gamma,g_0,\Lambda)$ for finite~$\Lambda$.

From now on we consider $d=4$, and assume all fields to be massive.
In an expansion of $S^0_{QFT}(\gamma,g_0,\Lambda)$ up to quadratic order in derivatives acting on $\gamma_{\mu\nu}$, $g_0$ and $\Lambda$, $S^0_{QFT}$ contains terms of the form
\begin{align}\label{w0}
S^0_{QFT}&(\gamma,g_0,\Lambda)=
\int d^4x \sqrt{\gamma}\Bigg( V^{0}(g_0,{\Lambda}) -L^{0}(g_0,{\Lambda}) R_4\nn \\
&+\gamma^{\mu\nu}\left\{\frac{1}{2}K^{0}_{ab}(g_0,{\Lambda})\partial_\mu g_0^a\partial_\nu g_0^b+K^{0}_{a\lambda}(g_0,{\Lambda})\partial_\mu g_0^a\partial_\nu \Lambda +\frac{1}{2} K^{0}_{\lambda\lambda}(g_0,{\Lambda})\partial_\mu \Lambda\partial_\nu \Lambda
\right\} \Bigg) 
\end{align}
where $R_4$ denotes the curvature scalar constructed from $\gamma_{\mu\nu}$.
The origin of derivatives $\partial_\mu \Lambda(x)$ are vertices involving $\Lambda(x)$ in the field theory action in the kinetic action $S_k$ via the cutoff functions, and cannot be dropped for the computation of $S^0_{QFT}$ since invariance under local Weyl variations $\sim \sigma(x)$ would not hold.

On the other hand, in terms of $\hat g_0^a$ and  $\hat \gamma_{\mu\nu}$, $\hat S_{QFT}(\hat\gamma, \hat g_0)$ satisfies
\beq\label{hatSQFT}
\Lambda(x)\frac{\delta}{\delta \Lambda(x)} \hat S_{QFT}(\hat\gamma, \hat g_0)=0\; ,
\eeq
i.e. $\hat S_{QFT}(\hat\gamma, \hat g_0)$ does not depend explicitely on $\Lambda(x)$ or its derivatives.
For a constant cutoff eq.~\eqref{hatSQFT} corresponds just to naive power counting involving the dimensionful rescaled metric $\hat\gamma$ and derivatives. However, the absence of derivatives acting on $\Lambda$ is not trivial.
The expansion of $\hat S_{QFT}(\hat\gamma, \hat g_0)$ to second order in derivatives reads
\begin{align}\label{w0h}
\hat S_{QFT}(\hat\gamma, \hat g_0)=
\int d^4x \sqrt{\hat \gamma}\Big( \hat V(\hat g_0) -\hat L(\hat g_0) R_4(\hat \gamma)
+\hat \gamma^{\mu\nu}\frac{1}{2}\hat K_{ab}(\hat g_0)\partial_\mu \hat g_0^a\partial_\nu \hat g_0^b 
\Big) \; . 
\end{align}
It is straightforward to replace $\hat\gamma$ by $\gamma$ and $\hat g^a_0$ by $g^a_0$ in $\hat S_{QFT}(\hat\gamma, \hat g_0)$. This generates terms of the form $K^{0}_{a\lambda}\partial_\mu g_0^a\partial_\nu \Lambda$ and $K^{0}_{\lambda\lambda}\partial_\mu \Lambda \partial_\nu \Lambda$, and a $\Lambda$ dependence of $V^0$, $L^0$ and all $K^0$. Given the local eq.~\eqref{localRGEW0}, these terms have to match the ones in $S^0_{QFT}(\gamma, g_0, \Lambda)$.

Next we introduce transformations characterized by a local parameter $\tau(x)$ which describe the dependence of $\hat\gamma$ and $\hat g^a_0$ on the local cutoff $\Lambda(x)$ (and which differ from Weyl transformations since $\gamma_{\mu\nu}$ remains inert):
\beq\label{dtau}
\delta_\tau \Lambda = \tau \Lambda\; ,\quad
\delta_\tau \hat\gamma_{\mu\nu} = 2\tau \hat\gamma_{\mu\nu}\; ,\quad 
\delta_\tau \hat g_0^a = \tau \beta^a_\tau\; ,\quad
\beta^a_\tau= \Lambda \frac{\delta \hat g_0^a}{\delta \Lambda}\; .
\eeq
For rescaled relevant couplings $\hat g_0^a$, $\beta^a_\tau$ contains a tree level term $-d_0^a$ as well as possible derivative terms. Recursion relations ensure that $\beta^a_\tau$ can again be expressed in terms of bare couplings.

The $\tau$ variation of $\hat S_{QFT}(\hat\gamma, \hat g_0)$ is given by
\beq
\delta_\tau\hat  S_{QFT}(\hat\gamma, \hat g_0) = \left(\Delta_\tau^\gamma + \Delta_\tau^\beta \right)
\hat S_{QFT}(\hat\gamma, \hat g_0)
\eeq
where
\beq
\Delta_\tau^\gamma= 2\tau \hat\gamma_{\mu\nu}\frac{\delta}{\delta \hat\gamma_{\mu\nu}}\; ,\quad
\Delta_\tau^\beta=\tau \beta^a_\tau \frac{\delta}{\delta \hat g^a_0}\;  .
\eeq

As in the case of local Weyl transformations considered by JO, the fact that local $\tau$ transformations are Abelian imposes constraints on the coefficients of the vacuum partition function. (In \cite{Osborn:1989td,Jack:1990eb,Osborn:1991gm} local Weyl transformations are applied to the vacuum partition function as function of the renormalized couplings in dimensional regularization, and after adding counter terms ${S^{CT}}_{QFT}$.) In the present case these consistency conditions originate from
\beq\label{conscondSQ}
\left[(\Delta_{\tau'}^\gamma + \Delta_{\tau'}^{\beta}),(\Delta_\tau^\gamma + \Delta_\tau^{\beta})\right] \hat S_{QFT}(\hat\gamma, \hat g_0) \overset{!}{=} 0\; .
\eeq
Inserting $\hat S_{QFT}(\hat\gamma, \hat g_0)$ from \eqref{w0h} into \eqref{conscondSQ} one obtains from terms proportional to \newline
$\left(\tau'\partial_\mu \tau-\tau \partial_\mu \tau'\right)\nabla^\mu \hat g^a_0$:
\beq\label{gradflowQ}
6\,\partial_a \hat L - \beta_\tau^b \hat K_{ba} \overset{!}{=} 0
\eeq
where $\partial_a$ denotes $\partial/\partial \hat g^a_0$. 
Eq.~\eqref{gradflowQ} corresponds to a gradient flow for $\beta_\tau^b$, with $\hat K_{ba}$ a metric in the space of couplings. Within the considered order in derivatives this result holds only for the terms without derivatives in $\beta_\tau^b$.

Some comments on the cutoff dependences are in order: 
First, since $\hat L$ does not depend explicitely on $\Lambda$, eq.~\eqref{gradflowQ} implies
\beq\label{ltheorem}
\Lambda \frac{d}{d\Lambda}\hat L = \frac{1}{6} \beta_\tau^a \hat K_{ab} \beta_\tau^b,
\eeq
i.e. an ``l-theorem''
\beq\label{ltheorem1}
\Lambda \frac{d}{d\Lambda}\hat L > 0 \ \text{iff the metric}\ \hat K_{ab}\ \text{is positive.}
\eeq

Second, the coefficients $ \hat V(\hat g_0)$, $\hat L(\hat g_0)$ and $\hat K_{ab}(\hat g_0)$ in $\hat S_{QFT}$ are all dimensionless.
After the replacement of $\hat\gamma$ and $\hat g_0$ as in eqs.~\eqref{rescgamma} and \eqref{rescg},
$\sqrt{\hat\gamma}$ scales as $\Lambda^4$. Rescaled bare masses $\hat m_0^2$ scale as $m_0^2/\Lambda^2$. Subsequently the coefficients $ V^{0}(g_0,{\Lambda})$, $L^{0}(g_0,{\Lambda})$ and $K^{0}_{ab}(g_0,{\Lambda})$ in $S^0_{QFT}$ assume their canonical dimensions which determine their leading powers in $\Lambda$ modified by powers of $\ln(\hat m_0^2) = \ln(m_0^2/\Lambda^2)$. These divergences in $V^0$, $L^0$ and $K^0$ include both subdivergences and superficial divergences, i.e. the information on subdivergences is present via the $\Lambda$ dependence of these coefficients. This information can be extracted once the beta functions are computed via the consistency condition \eqref{gradflowQ}. The beta functions for relevant couplings like masses are cutoff dependent in general. Particular conditions on the UV limit of couplings including relevant couplings do not follow from the present formalism.

\section{Towards a Holograpic Dual}

In the first part of this Section we review the construction of a 4d action on the boundary $\partial{\cal M}$ of a 5d bulk ${\cal M}$ following, to a large extend, the Hamilton-Jacobi approach used in \cite{deBoer:1999tgo,Fukuma:2002sb}.
With ans\"atze for the 5d Lagrangian ${\cal L}_5$ and the action $\bar{S}$ on the boundary, both expanded to second order in derivatives, we establish relations between the corresponding parameters which follow from the Hamiltonian constraint ${\cal H}_5=0$.

For the metric in the bulk we take
\beq\label{ds3}
ds^2=\frac{l^2}{4}\frac{d\rho^2}{\rho^2} + {G}_{4ij}(x,\rho)dx^i dx^j\; .
\eeq
From now on indices $\mu,\nu$ refer to indices in 5 dimensions. The 5d metric will be denoted by $G_{5\mu\nu}$, and $i,j$ will be indices in 4 dimensions. The 5d Lagrangian ${\cal L}_5$ includes scalars $\Phi^a$ and reads, expanded to second order in derivatives\footnote{A non-Einstein factor Z in front of $R_5$ can be removed by a Weyl rescaling of $G_5$ and a subsequent diffeomorphism such that no additional degree of freedom in the final boundary action remains.},
\beq\label{L51}
{\cal L}_5=-R_5 +\frac{1}{2}G_5^{\mu\nu}L_{ab}(\Phi)\partial_\mu \Phi^a \partial_\nu \Phi^b +V(\Phi)\; .
\eeq

Next we consider the 5d action regularized for $\rho \to 0$ by $\rho \geqq \bar{\rho}$,
\beq\label{sgrav3}
2\kappa_5^2 S=\int d^5x \sqrt{G_5} {\cal L}_5 + S_{GH}
=\int d^4x \sqrt{{G_4}} \int_{\bar \rho} d\rho  \frac{l}{2\rho}  {\cal L}_5 + S_{GH}\; ,
\eeq
where $S_{GH}$ denotes the Gibbons-Hawking action on the boundary $\partial{\cal M}$ at $\rho=\bar{\rho}$: 
\beq
S_{GH}=-\frac{2\bar \rho}{l}\int d^4x \sqrt{\bar{g}} \bar{g}^{ij}\dot{\bar{g}}_{ij}
\eeq
where $\bar{g}_{ij}$ denotes ${G}_{4ij}(x,\rho=\bar{\rho})$. Here and below a dot denotes a derivative $\partial/\partial \rho$.

In the Hamilton-Jacobi formalism, the bulk fields ${G}_{4ij}$ and $\Phi^a$ are replaced by their solutions of the equations of motion with $\bar{g}_{ij}$ and $\bar{\phi}^a$ as boundary conditions on $\partial{\cal M}$ corresponding to $\rho=\bar{\rho}$. This allows to express $S$ as function of $\bar{g}_{ij}$ and $\bar{\phi}^a$, denoted by $\bar{S}(\bar{g}_{ij},\bar{\phi}^a)$. For completeness, details of this procedure are given in Appendix~A. Hamiltonian constraints will relate $\bar{S}(\bar{g}_{ij},\bar{\phi}^a)$ to $\bar{{\cal L}}_5$, the bulk Lagrangian at $\rho=\bar{\rho}$ as function of $\bar{g}_{ij}$ and $\bar{\phi}^a$. 

For $\bar{S}$ expanded to second order in derivatives we make the ansatz ($g_{ij}$ and $\phi^a$ denote $\bar g_{ij}$ and $\bar\phi^a$ to lowest order in derivatives, respectively)
\beq\label{sregmain}
\bar{S}=\int d^4x \sqrt{{g}}\left(W({\phi}) - L({\phi}) R_4({g})
+\frac{1}{2}K_{ab}({\phi})\nabla^i{\phi}^a\nabla_i{\phi}^b 
\right)\; .
\eeq
The flow equations for the leading terms of ${g}_{ij}$ and ${\phi}^a$ in an expansion in derivatives become (see Appendix~A and \cite{deBoer:1999tgo})
\begin{align}\label{flow}
\dot{{g}}_{ij}&=\frac{l}{6\bar\rho} g_{ij} W \; ,\\
\label{flow1}
\dot{{\phi}}^a&=-\frac{l}{2\bar\rho}L^{ab} \partial_b W
\; .
\end{align}

The Hamiltonian constraint ${\cal H}_5=0$ imposes the following relations between the coefficients in $\bar{{\cal L}}_5$ and $\bar{S}$ which are derived in Appendix~A:
\begin{align}
\label{Happm1}
V=& -\frac{1}{3} W^2+\frac{1}{2} \partial_a W L^{ab} \partial_b W\; ,\\
\label{Happm2}
0=&1+\frac{1}{3} W L- \partial_a W L^{ab}\partial_b L\; ,\\
\label{Happm3}
0=&L_{ab} + W \left(\frac{1}{3} K_{ab} +2\partial_a \partial_b L\right)
- \partial_c W L^{cd}(\partial_d K_{ab}-\partial_a K_{bd}-\partial_b K_{ad})\; ,\\
\label{Happm4}
0=&W \partial_a L + \partial_b W L^{bc} K_{ca}\; .
\end{align}
These agree with \cite{deBoer:1999tgo,Fukuma:2002sb}.

We write $g_{ij}$ in the form
\beq\label{ansg}
g_{ij}=a({\phi},\bar{\rho})\tilde{\gamma}_{ij}
\eeq
with $\tilde{\gamma}_{ij}$ independent from $\bar{\rho}$.
From eq.~\eqref{flow} the flow equation for $a({\phi},\bar{\rho})$ reads
\beq
\dot{a}= a \frac{l}{6\bar \rho} W\; .
\eeq

Next we consider local transformations by $\tau(x)$.
To $g_{ij}$ resp. $a$ we attribute $\tau$ transformation laws
\beq\label{dscaleg}
\delta_\tau g_{ij} = 2\tau g_{ij}\; , \quad \delta_\tau a = 2\tau a
\eeq
such that we can identify $g_{ij}$ with the QFT metric $\hat \gamma_{ij}$, and $a$ to the QFT cutoff $\Lambda^2$ which varies under Weyl transformations as in eq.~\eqref{eq:weylresc}.

Let us recall the role of the parameter $\bar\rho$ for the boundary action $\bar S$: $\bar \rho$ does not appear explicitely in $\bar S$, only implicitely in $\phi(\bar\rho)$ and $a({\phi},\bar{\rho})$. As argued in \cite{deBoer:1999tgo} one can perform a change of variables and consider $a$ instead of $\bar\rho$ as independent variable. 
The flow equation for $\phi(a)$ reads now
\beq
a\frac{\delta {\phi}^a}{\delta a}=-\frac{3}{W} L^{ab}\partial_b W\; .
\eeq

For $\tau$ transformations of $\phi$ we assume that these are induced by the dependence of $\phi$ on $a$:
\beq\label{betaa}
\delta \phi^a = \tau \beta_\tau^a\; ,\quad 
\beta_\tau^a= 2 a\frac{\delta {\phi}^a}{\delta a}=-\frac{6}{W} L^{ab}\partial_b W \; .
\eeq
For the local $\tau$ transformations of $\bar S(g,\phi)$ one obtains
\beq\label{weylS}
\delta_\tau \bar S(g,\phi) = (\Delta_\tau^g + \Delta_\tau^{\beta})\bar S(g,\phi)
\eeq
with
\beq
\Delta_\tau^g=2\tau g_{ij}\frac{\delta}{\delta g_{ij}}\; ,\quad 
\Delta_\tau^\beta=\tau \beta_\tau^a \frac{\delta}{\delta \phi^a}\; .
\eeq
As in the case of $\hat S_{QFT}(\hat\gamma, \hat g_0)$, the local $\tau$ transformations of $\bar S(g,\phi)$ in \eqref{weylS} imply consistency conditions like in \eqref{gradflowQ}. In terms of the coefficients of $\bar S$ they read
\beq\label{gradflowB}
6\,\partial_a  L - \beta_\tau^b K_{ba} \overset{!}{=} 0
\eeq

With $\beta_\tau^a$ from \eqref{betaa} eq.~\eqref{gradflowB} becomes
\beq\label{conscondS1}
W\partial_a L + K_{ab} L^{bc}\partial_c W \overset{!}{=} 0\; .
\eeq
Eq. \eqref{conscondS1} corresponds to the Hamiltonian constraint \eqref{Happm4}, which will play an important role in the next Section.

\section{Matching the Bulk to the QFT}

The starting point of most explicit applications of the holographic renormalization group is a 5d Lagrangian ${\cal L}_5$, and resulting properties of the effective action and constraints on $\beta$ functions of a corresponding 4d QFT. Here we ask the opposite question: Given a set of $\beta$ functions describing the running of bare (marginal or relevant) couplings $g_0^a$ of a 4d QFT, can this RG flow always be related to the flow of fields $\phi^a$ living on the boundary of a 5d manifold? Or, can the vacuum partition function of a 4d QFT in a gravitational background as function of local bare couplings $g_0^a(x)$ always be related to an action for fields $\phi^a$ on the boundary of a 5d manifold?

As a first step in this direction we consider both functionals to first nontrivial order in an expansion in 4d derivatives. To this order the vacuum partition function $\hat S_{QFT}$ of a 4d QFT as function of the rescaled metric $\hat \gamma$ and rescaled local couplings $\hat g_0$ is specified by the three coefficients $\hat V(\hat g_0)$, $\hat L(\hat g_0)$ and $\hat K_{ab}(\hat g_0)$ in eq. \eqref{w0h}. In turn, the boundary action $\bar S$ as function boundary fields $\phi$ is specified by three coefficients $W(\phi)$, $L(\phi)$ and $K_{ab}(\phi)$ in eq. \eqref{sregmain}. 

We assume that the rescaled QFT metric $\hat\gamma_{ij}$ can be identified with the boundary metric $g_{ij}$, the bare rescaled QFT couplings $\hat g_0^a$ with the boundary fields $\phi^a$, and the QFT cutoff $\Lambda^2$ with the coefficient $a$ in eq.~\eqref{ansg} (and not with the boundary coordinate $\bar\rho$ unless $lW=6$). Note that both $\hat\gamma_{ij}$ and $g_{ij}$ diverge identically for $\Lambda^2 = a \to \infty$.

Given the coefficients $\hat V(\hat g_0)$, $\hat L(\hat g_0)$ and $\hat K_{ab}(\hat g_0)$ in $\hat S_{QFT}$, a correspondence 
\begin{align} 
W(\bar\phi^a)&= \hat V(\hat g_0^a=\phi^a) \label{qftvsb}\\
L(\bar\phi^a)&= \hat L(\hat g_0^a=\phi^a)\\
K_{ab}(\bar\phi^a)&= \hat K_{ab}(\hat g_0^a=\phi^a)
\end{align}
faces the following potential obstruction: If interpreted as coefficients of a boundary action $\bar S$ resulting from the classical equations of motion based on a 5d Lagrangian ${\cal L}_5$ including gravity, $W$, $L$ and $K_{ab}$ must satisfy the Hamiltonian constraints \eqref{Happm1}-\eqref{Happm4}. 

In the present approach these have to be interpreted as equations which determine the coefficients $V(\Phi)$ and $L_{ab}(\Phi)$ in ${\cal L}_5$ in \eqref{L51}.
Eq.~\eqref{Happm1} determines $V(\Phi)$, and the Hamiltonian constraint~\eqref{Happm3} allows, in principle, to compute $L_{ab}(\Phi)$ in terms of $W$, $L$ and $K_{ab}$. For $n_g$ bare couplings, the Hamiltonian constraint~\eqref{Happm4} seems to impose $n_g$ additional conditions.
At this point the consistency conditions originating from local $\tau$ transformations -- satisfied by both $\hat S_{QFT}$ and $\bar S$ -- come into play: First, recall that the $\beta$ functions $\beta_\tau^a$ describe the flow of both $g_0^a$ in the QFT with the cutoff $\Lambda$, and of the boundary fields $\phi^a$ with $a=\Lambda^2$, respectively. Second, the consistency conditions imply the same gradient flows eq.~\eqref{gradflowQ} and eq.~\eqref{gradflowB}, hence the $\beta$ functions for the QFT and the boundary fields coincide. Third, the consistency condition in the form \eqref{conscondS1} implies the validity of the $n_g$ Hamiltonian constraints \eqref{Happm4} without further assumptions. Hence it remains only the Hamiltonian constraint~\eqref{Happm2} which, with $L_{ab}(\Phi)$ determined by \eqref{Happm3}, imposes one extra condition on the coefficients of $\hat S_{QFT}$. 

Finally a successful QFT/bulk correspondence implies another relation:
As shown in \cite{Freedman:1999gp}, the weaker energy condition (or null energy condition) in the 5d bulk implies a c-theorem, i.e. the existence of a c-function which decreases towards the infrared. Following \cite{Fukuma:2002sb} adapted to our conventions, this c-function is given by 
\beq\label{cfunct}
c(\phi)=-\frac{1}{W^3}
\eeq
and satisfies
\beq\label{ctheor}
2a\frac{d}{da} c(\phi) \equiv \beta_\tau^a \frac{d}{d\phi^a} c(\phi) \geq 0\; .
\eeq
With $c(\phi)$ as in eq.~\eqref{cfunct} and $\beta_\tau^a$ from eq.~\eqref{betaa}, the c-theorem is equivalent to
\beq
-\beta_\tau^a \frac{1}{W^3} L_{ab} \beta_\tau^b \geq 0\; .
\eeq
Accordingly the metric in field space
\beq
-\frac{1}{W^3} L_{ab}
\eeq
must be semi-positive. If this condition (and the Hamiltonian constraint (3.9)) are satisfied, the c-theorem applies also to the 4d QFT, but with respect to the Wilsonian UV cutoff $\Lambda$: Using the relation \eqref{qftvsb}, the expression
\beq
c_{QFT}(\hat{g}_0)= -\frac{1}{\hat{V}^3(\hat{g}_0)}
\eeq
increases for increasing $\Lambda$, i.e. $\hat{V}(\hat{g}_0)$ increases for increasing $\Lambda$ due to the running of $\hat{g}_0^a$ with $\Lambda$.

We recall that the origin of the Hamiltonian constraint in general relativity can be traced back to the (explicit) independence of the boundary action $\bar S$ on the coordinate of the boundary. The corresponding feature of the QFT action in the form of $\hat S_{QFT}$ is that $\hat S_{QFT}$ satisfies eq.~\eqref{hatSQFT} implying its (explicit) independence on the cutoff $\Lambda$.
These are the central results of the present approach.

\section{Summary and Outlook}

The introduction of a local ultraviolet cutoff in the kinetic terms of a generic (here: multi-flavour scalar) QFT together with local couplings in a curved background allows to gain new insight into properties of the bare vacuum partition function as function of bare couplings. 

First, it allows to realize local Weyl transformations under which the local cutoff transforms; here the local component of the cutoff plays the role of a compensator allowing for a local RGE satisfied by the bare vacuum partition function. This local RGE can be solved in terms of a rescaled space-time metric, and rescaled bare couplings such that they are dimensionless. (For marginal bare couplings, this rescaling is trivial.) Next one can study the variation of the bare vacuum partition function as function of the rescaled space-time metric and rescaled bare couplings under local variations of the cutoff. Since these variations are Abelian they imply consistency conditions similar to the Weyl consistency conditions introduced by JO.
In an expansion to first nontrivial order in derivatives we studied the consequences of these consistency conditions and found that they imply a gradient flow for the beta functions describing the cutoff dependence of the rescaled bare couplings.

Second, we reviewed the holographic description of the renormalization group flow which implies Hamiltonian constraints on the action on the boundary of a 5d manifold, following to a large extend \cite{deBoer:1999tgo}. At first sight the number of these constraints does not allow to identify an arbitrary 4d vacuum partition function of local bare couplings with an action on the boundary of a 5d manifold, identifying scalar fields on the boundary with local rescaled bare couplings in the 4d QFT. It turns out, however, that the consistency condition to be satisfied by coefficients of the 4d vacuum partition function correspond precisely to one of the Hamiltonian constraints such that only one additional condition on the 4d vacuum partition function remains. (An open question is whether a particular choice of the cutoff function of the 4d QFT allows to satisfy this remaining condition.)

We underline that this consistency condition coincides with the one of the QFT only if the holographic beta functions describe the flow of bulk fields with the field dependent scale factor $a(\phi,\bar{\rho})$ multiplying the induced metric on the boundary and not, as it is frequently assumed, with $\bar{\rho}$ directly. This is plausible, given that fields in the bulk modify the metric in the bulk via the Einstein equations, and physical distances in the bulk depend on the metric in the bulk.

Although this result was obtained only to first nontrivial order in an expansion in derivatives it nourishes the hope that generic non-conformal quantum field theories may allow for a description in terms of fields on the boundary of a 5d bulk subject to equations of motion including general relativity. Of course, to establish such a correspondence beyond an expansion in derivatives is a more ambitious task.

Independently thereof, the present approach may also be helpful for the study of the ultraviolet behaviour of 4d QFTs such as the search for UV fix points including relevant couplings, related to extrema of the coefficient $\hat L^0(\hat g_0)$ in \eqref{w0h}. The generalisation to QFTs including Fermions and/or gauge fields remains a task for the future.

\appendix

\section{Appendix}

\section*{The Hamiltonian Constraint}

In this Appendix we sketch the Hamilton-Jacobi formalism relating an action $\bar{S}$, a functional of fields $\bar{g},\bar{\phi}$ on the boundary $\partial {\cal M}$, to the bulk Lagrangian ${\cal L}_5$ evaluated on the boundary. It leads to flow equations for the fields $\bar{g},\bar{\phi}$ with respect to the extra coordinate. The Hamiltonian constraint implies relations between the parameters of $\bar{S}$ and the bulk Lagrangian ${\cal L}_5$. We assume ${\cal L}_5$ quadratic in derivatives.

For the bulk metric we take
\beq
ds^2=\frac{l^2}{4}\frac{d\rho^2}{\rho^2} +{G}_{4ij}dx^i dx^j\; ,\qquad \sqrt{G_5}= b\sqrt{{G_4}}\; ,\qquad b=\frac{l}{2\rho}\; .
\eeq
For ${\cal L}_5$ we take
\beq\label{L5}
{\cal L}_5= -R_5 +V(\Phi) +\frac{1}{2} G_5^{\mu\nu} L_{ab}(\Phi)\partial_\mu\Phi^a \partial_\nu\Phi^b\; .
\eeq
The regularized bulk action is
\beq\label{sreg0}
2\kappa_5^2 S= 
\int d^5x \sqrt{G_5} {\cal L}_5 + S_{GH}
= \int d^4x \int^\infty_{\bar{\rho}} d\rho \sqrt{{G_4}} b {\cal L}_5 + S_{GH} 
\eeq
where $S_{GH}$ denotes the Gibbons-Hawking action on the boundary $\bar{\rho}$:
\beq
S_{GH}=-\frac{1}{\bar{b}}\int d^4x \sqrt{{g}} {g}^{ij}\dot{{g}}_{ij}
\eeq
where a dot denotes $\partial/\partial \bar{\rho}$ and $\bar{b}=\frac{l}{2\bar{\rho}}$.

Solving the equations of motion for ${G}_{4ij}$ and $\Phi$, $S$ becomes a function of the corresponding values ${g},{\phi}$ on the boundary $\bar{\rho}$ and will be denoted by $\bar{S}({g},{\phi})$.
${\cal L}_5$ evaluated on the boundary will be denoted by $\bar{{\cal L}}_5$. 
Separating 4d derivatives from derivatives w.r.t. $\rho$ and after a partial integration w.r.t. $\rho$, $\bar{{\cal L}}_5$ becomes
\beq\label{L5app}
\bar{{\cal L}}_5
=-R_4({g})+\frac{1}{4\bar{b}^2}(\dot{{g}}_{ij}{g}^{ik}{g}^{jl}\dot{{g}}_{kl} -({g}^{ij}\dot{{g}}_{ij})^2)
+\frac{1}{2}L_{ab}({\phi})\big(\frac{1}{\bar{b}^2}\dot{{\phi}}^a{\dot{\phi}}^b+{g}^{ij}\partial_i{\phi}^a\partial_j{\phi}^b\big)+V({\phi})\; .
\eeq
In $\bar{{\cal L}}_5$, contributions from partial integration of second derivatives w.r.t. $\rho$ have been cancelled by the Gibbons-Hawking action.

In order to relate $\bar{S}({g},{\phi})$ to $\bar{{\cal L}}_5({g},{\phi})$ via the solutions of the equations of motion for ${G}_{4ij}$ and $\Phi^a$ inserted into $\sqrt{G_5}{\cal L}_5$
we combine the bulk fields ${G}_{4ij}$ and $\Phi^a$ into fields $q_\alpha$ were $\alpha$ denote the corresponding indices:
\beq
q_\alpha=\left\{{G}_{4ij}, \Phi^a\right\}\; .
\eeq
It is convenient to introduce
\beq
\tilde{\cal L}_5=\sqrt{G_5}{\cal L}_5= b\sqrt{{G_4}}{\cal L}_5\; .
\eeq
The equations of motion are
\beq
\frac{\delta \tilde{\cal L}_5}{\delta q_\alpha}-\partial_i \frac{\delta \tilde{\cal L}_5}{\delta \partial_iq_\alpha}-\partial_\rho \frac{\delta \tilde{\cal L}_5}{\delta  \dot{q}_\alpha} =0\; . 
\eeq
The variations of $S$ with $\delta q_\alpha$ and its derivatives are
\beq
\frac{2\kappa_5^2}{\sqrt{g}} \delta S = \int d^4x \int_{\bar{\rho}(x)}^\infty d\rho \left( \frac{\delta \tilde{\cal L}_5}{\delta q_\alpha} \delta q_\alpha + \frac{\delta \tilde{\cal L}_5}{\delta \partial_i q_\alpha} \delta \partial_i q_\alpha + \frac{\delta \tilde{\cal L}_5}{\delta \dot{q}_\alpha} \delta \dot{q}_\alpha \right)\; .
\eeq
Using the equations of motion and partial integration under the $d\rho$ integral one obtains
\beq
\frac{2\kappa_5^2}{\sqrt{g}}\delta S =\left[\int d^4x \frac{\delta \tilde{\cal L}_5}{\delta \dot{q}_\alpha} \delta q_\alpha\right]_\infty
-\left[\int d^4x \frac{\delta \tilde{\cal L}_5}{\delta \dot{q}_\alpha} \delta q_\alpha\right]_{\partial{\cal M}}\; .
\eeq
With $\bar{q}_\alpha= q_{\alpha}|_{\partial{\cal M}}$ and $\bar{S}={2\kappa_5^2} S(q_\alpha=\bar{q}_\alpha)$ one obtains
\beq
\frac{1}{\sqrt{g}}
\frac{\delta \bar{S}(\bar q)}{\delta \bar{q}_\alpha}
=-\frac{1}{\sqrt{{g}}} \frac{\delta \tilde{\cal L}_5}{\delta \dot{\bar{q}}_\alpha}
=-\bar b \frac{\delta \bar{\cal L}_5}{\delta \dot{\bar{q}}_\alpha}
\eeq
or, returning from $\bar{q}_\alpha$ to ${g}_{ij},{\bar{\phi}^a}$,
\beq\label{dSdq}
\frac{\delta \bar{\cal L}_5}{\delta \dot{{g}}_{ij}}
= -\frac{1}{\bar b\sqrt{{g}}} \frac{\delta \bar{S}}{\delta {g}_{ij}}\; ,\qquad
\frac{\delta \bar{\cal L}_5}{\delta \dot{{\phi}}^a}
= -\frac{1}{\bar b\sqrt{{g}}} \frac{\delta \bar{S}}{\delta {\phi}^a}\; .
\eeq
(Conventionally $\bar b\frac{\delta \bar{\cal L}_5}{\delta \dot{{g}}_{ij}}$ and $\bar b\frac{\delta \bar{\cal L}_5}{\delta \dot{{\phi}}^a}$ are denoted by canonical momentum variables $\pi_{ij}$ and $\pi_a$, respectively \cite{deBoer:1999tgo,Fukuma:2002sb}.)

With ${\cal L}_5$ and hence $\bar{\cal L}_5$ quadratic in dotted derivatives it is straightforward to express $\frac{\delta \bar{\cal L}_5}{\delta \dot{{g}}_{ij}}$ and $\frac{\delta \bar{\cal L}_5}{\delta \dot{{\phi}}^a}$ in \eqref{dSdq} in terms of $\dot{{g}}_{ij}$ and $\dot{{\phi}}^a$, and to obtain flow equations for $\dot{{g}}_{ij}$ and $\dot{{\phi}}^a$:
\begin{align}\label{flowa1}
\dot{{g}}_{ij}&=-2\bar{b}\left({g}_{ik}{g}_{jl}-\frac{1}{3}{g}_{ij}{g}_{kl}\right) 
\frac{1}{\sqrt{{g}}} \frac{\delta \bar{S}}{\delta {g}_{kl}}
\; ,\\
\label{flowa2}
\dot{{\phi}}^a&=-\bar{b}L^{ab} \frac{1}{\sqrt{{g}}}\frac{\delta \bar{S}}{\delta {\phi}^b}
\; .
\end{align}

The Hamilton density ${\cal H}_5$ is given by
\beq\label{H5}
{\cal H}_5=-\dot{g}_{ij}\frac{\delta {\cal L}_5}{\delta \dot{g}_{ij}} - \dot{\Phi}^a \frac{\delta {\cal L}_5}{\delta \dot{\Phi}^a}+{\cal L}_5\; .
\eeq
Replacing the bulk fields by the solutions of the equations of motion, ${\cal H}_5$ satisfies the Hamiltonian constraint\footnote{Without gauge fixing of the bulk metric, i.e. introducing lapse and shift as in the ADM decomposition, the Hamiltonian constraint follows from the variation w.r.t. the lapse.}
\beq\label{Hconstraint}
{\cal H}_5=0
\eeq
which (evaluated on the boundary $\rho=\bar{\rho}$) is equivalent to $\partial \bar{S}/\partial \bar{\rho}=0$ corresponding to the only dependence of $\bar{S}$ on the
coordinate $\bar{\rho}$ via arguments of the metric ${g(\bar{\rho})}$ and the fields ${\phi}(\bar{\rho})$.
In terms of $\bar{S}$, the Hamiltonian constraint ${\cal H}_5=0$ evaluated on the boundary becomes
\beq\label{HS1}
\frac{1}{\sqrt{{g}}}\frac{\delta \bar{S}}{\delta {g}_{ij}}
\left({g}_{ik}{g}_{jl}-\frac{1}{3}{g}_{ij}{g}_{kl}\right)
\frac{1}{\sqrt{{g}}}\frac{\delta \bar{S}}{\delta {g}_{kl}}
+\frac{1}{2}\frac{1}{\sqrt{{g}}} \frac{\delta \bar{S}}{\delta {\phi}^a} L^{ab} 
\frac{1}{\sqrt{{g}}} \frac{\delta \bar{S}}{\delta {\phi}^b} ={\cal L}_4
\eeq
where
\beq\label{l4}
{\cal L}_4=-R_4+\frac{1}{2}L_{ab}({\phi}){g}^{ij}\,\partial_i {\phi}^a \partial_j {\phi}^b +V({\phi})\; .
\eeq
Eq.~\eqref{HS1} coincides with the Hamilton-Jacobi constraint in \cite{deBoer:1999tgo,Fukuma:2002sb}. 

Next we consider $\bar{S}$ expanded to second order in derivatives,
\beq\label{srega}
\bar{S}=\int d^4x \sqrt{{g}}\left(W({\phi}) - L({\phi}) R_4({g})
+\frac{1}{2}K_{ab}({\phi})\nabla^i{\phi}^a\nabla_i{\phi}^b 
\right)\; .
\eeq
First, following \cite{deBoer:1999tgo}, we expand the flow equations \eqref{flowa1} and \eqref{flowa2} in derivatives.
To lowest order one obtains
\begin{align}\label{flow2}
\dot{{g}}_{ij}&=\frac{1}{3} g_{ij} \bar{b} W = \frac{l}{6\bar \rho} W g_{ij} \; ,\\
\dot{{\phi}}^a&=-\bar{b}L^{ab} \partial_b W =-\frac{l}{2\bar\rho}L^{ab} \partial_b W
\; .
\end{align}
(In a slight abuse of notation, the solutions $\hat g_{ij}$ and $\hat \phi^a$ of the flow equations \eqref{flowa1} and \eqref{flowa2} can be expanded as $\hat g_{ij}=g_{ij}+ \text{derivatives}$, $\hat \phi^a=\phi^a+ \text{derivatives}$. Inserted into $\bar{S}(\hat g_{ij},\hat \phi^a)$ and expanding $\bar{S}$ as in \eqref{srega}, the derivative terms in $\hat g_{ij}$ and $\hat \phi^a$ induce flows for the coefficients $L$ and $K_{ab}$ which have to be consistent with the Hamiltonian constraints below.)

Using ${\cal L}_4$ from \eqref{l4} the Hamiltonian constraint \eqref{HS1} implies relations between the coefficients in $\bar{S}$ and ${\cal L}_4$.
From the terms without derivatives $\partial_i$ on both sides of \eqref{HS1} one obtains
\beq\label{Happ1}
V = -\frac{1}{3}
W^2
+\frac{1}{2} \partial_a W L^{ab} \partial_b W\; .
\eeq
From terms $\sim R_4$ one obtains
\beq\label{Happ2}
1+\frac{1}{3} W L- \partial_a W L^{ab}\partial_b L=0\; .
\eeq
From terms $\sim \nabla^i {\phi}^a \nabla_i {\phi}^b$ one obtains
\beq\label{Happ3}
L_{ab} + W \left(\frac{1}{3} K_{ab} +2\partial_a \partial_b L\right)
- \partial_c W L^{cd}(\partial_d K_{ab}-\partial_a K_{bd}-\partial_b K_{ad})
=0\; .
\eeq
Finally terms $\sim \nabla^i \nabla_i {\phi}^a$ require
\beq\label{Happ4}
W \partial_a L + \partial_b W L^{bc} K_{ca}=0\; .
\eeq

\end{document}